\documentclass[journal]{IEEEtran}
\usepackage[utf8]{inputenc}
\usepackage{newunicodechar}
\newunicodechar{（}{(}
\newunicodechar{）}{)}
\usepackage{cite}
\usepackage{balance} % 在导言区添加

\ifCLASSINFOpdf
  \usepackage[pdftex]{graphicx}
\else
\fi
\usepackage[detect-all]{siunitx}
\usepackage{amsmath}
\usepackage{amssymb}
\usepackage{subcaption} 

\usepackage{booktabs}
\usepackage{siunitx}
\usepackage{amsthm}
\theoremstyle{remark}
\newtheorem*{remark}{Remark}
\usepackage{placeins} % 提供 \FloatBarrier
\usepackage{svg}          % 需要 Inkscape 支持
\begin{document}
\setlength{\parindent}{0pt}
%
% paper title
% Titles are generally capitalized except for words such as a, an, and, as,
% at, but, by, for, in, nor, of, on, or, the, to and up, which are usually
% not capitalized unless they are the first or last word of the title.
% Linebreaks \\ can be used within to get better formatting as desired.
% Do not put math or special symbols in the title.
\title{A Robust CSI-Based Scatterer Geometric Reconstruction Method for 6G ISAC System}
%
%
% author names and IEEE memberships
% note positions of commas and nonbreaking spaces ( ~ ) LaTeX will not break
% a structure at a ~ so this keeps an author's name from being broken across
% two lines.
% use \thanks{} to gain access to the first footnote area
% a separate \thanks must be used for each paragraph as LaTeX2e's \thanks
% was not built to handle multiple paragraphs
%

\author{Yubin Luo,
        Li Yu,
        Tao Wu, Yuxiang Zhang,
        and~Jianhua Zhang,~\IEEEmembership{Senior Member,~IEEE}% <-this % stops a space
\thanks{Corresponding
author: Li Yu. Yubin Luo, Li Yu, Tao Wu, Yuxiang Zhang and Jianhua Zhang are with the State Key Laboratory of Networking and Switching
Technology, Beijing University of Posts and Telecommunications, Beijing
100876, China. (e-mail: rx105ksai@gmail.com; li.yu@bupt.edu.cn; vvt@bupt.edu.cn; zhangyx@bupt.edu.cn;
jhzhang@bupt.edu.cn;)
  }% <-this % stops a space
}

% note the % following the last \IEEEmembership and also \thanks - 
% these prevent an unwanted space from occurring between the last author name
% and the end of the author line. i.e., if you had this:
% 
% \author{....lastname \thanks{...} \thanks{...} }
%                     ^------------^------------^----Do not want these spaces!
%
% a space would be appended to the last name and could cause every name on that
% line to be shifted left slightly. This is one of those "LaTeX things". For
% instance, "\textbf{A} \textbf{B}" will typeset as "A B" not "AB". To get
% "AB" then you have to do: "\textbf{A}\textbf{B}"
% \thanks is no different in this regard, so shield the last } of each \thanks
% that ends a line with a % and do not let a space in before the next \thanks.
% Spaces after \IEEEmembership other than the last one are OK (and needed) as
% you are supposed to have spaces between the names. For what it is worth,
% this is a minor point as most people would not even notice if the said evil
% space somehow managed to creep in.

% The paper headers
\markboth{Journal of \LaTeX\ Class Files,~Vol.~14, No.~8, August~2015}%
{Shell \MakeLowercase{\textit{et al.}}: Bare Demo of IEEEtran.cls for IEEE Journals}

\maketitle

% As a general rule, do not put math, special symbols or citations
% in the abstract or keywords.
\hspace{1em}\begin{abstract}

Digital twin (DT) is envisioned as a core enabler of sixth generation (6G) mobile systems. As the prerequisite of DT, scatterer geometric reconstruction (SGR) in propagation environment is essential, which typically requires additional sensing equipment such as camera and LiDAR. With the 6G integrated sensing and communication (ISAC) evolving, we reinterpret linear sampling method (LSM) from a wireless-channel perspective and propose a CSI-based variant for sensor-free SGR: by exploiting shared channel characteristic of multipath and scattering, dual-use in-band CSI replaces scattered-field measurements typically required for solving LSM. However in practice, aperture-limited arrays degrade LSM’s robustness. To counter this, we propose matched-filtering-enhanced multi-frequency LSM (MF-MLSM). Multi-frequency data increases frequency-domain diversity, and matched filtering coherently aligns inter-frequency phases to avoid artifacts, both of which improve robustness. Experiments with limited apertures of 93.6$^\circ$, 144$^\circ$, 180$^\circ$ and SNRs of 27 and 12~dB demonstrate robust SGR with our approach.

\end{abstract}

% Note that keywords are not normally used for peerreview papers.

\IEEEpeerreviewmaketitle

\section{Introduction}
\IEEEPARstart Digital twin (DT) has emerged as central aspirations of sixth generation (6G) mobile communication systems, aiming to provide a deeper understanding of the environment and to support more informed and efficient operation across diverse scenarios~\cite{b1,b2}. In turn, this enables real-time situational awareness and predictive optimization of wireless network operation and control \cite{b3}. The effective construction of DT is inseparable from accurate scatterer geometric reconstruction (SGR), particularly the precise characterization of their spatial distribution and geometric shapes~\cite{b4,b5}.

\hspace{1em}Traditional SGR method can generally be divided into hardware-based measurements and image-processing approaches. The former is exemplified by omnidirectional LiDAR, which performs SGR by emitting laser pulses and measuring the round-trip time of reflected signals~\cite{b6}. The latter category is represented by multi-modal based image processing, which integrates data from multiple sensors, i.e., depth cameras. Scene geometry and object shapes can be recovered via techniques such as stereo vision~\cite{b7,b8}. In 6G wireless systems, however, these approaches require additional sensing modules that are decoupled from the signal-transmission infrastructure. At scale, such out-of-band instrumentation incurs substantial costs in deployment, calibration, and maintenance expenses that grow quickly in large, complex scenarios.

\hspace{1em}To support 6G DT without adding sensors, we adopt the integrated sensing and communication (ISAC) technology, which reuses the in-band communication data stream for sensing. One practical approach that makes SGR possible using only in-band ISAC pilot signals is electromagnetic inverse imaging (EII). EII recovers the geometry and electromagnetic properties of unknown objects by solving wave equations from far-field scattered field measurements \cite{b9}, and has proven effective in medical imaging, remote sensing, and geophysical exploration \cite{b10,b11}. Moreover, within ISAC, the signals themselves can replace explicit scattered-field measurements in EII. In particular, routine pilot signaling yields channel state information (CSI) that can serve as a surrogate for scattered field data: in the frequency domain, the standard multipath channel model is a reduced Green’s function representation of Maxwell wave physics, so the received baseband signals essentially superpose the same scattered fields \cite{b12,b13}. With basic calibration or background subtraction to isolate the scattered component, CSI can be fed directly to an EII solver, providing a communication native path to SGR for DT.

\hspace{1em}On this basis, within our ISAC framework we further mitigate computational burden of SGR using a low-complexity EII algorithm—the linear sampling method (LSM). Unlike iterative schemes, LSM is non-iterative, requiring only a single linear solve to reconstruct scatterer geometry \cite{b14}, thus minimizing runtime and memory. However, traditional LSM assumes a full $360^\circ$ arrangement of transmit and receive antennas around the target, which is often impractical in operational ISAC deployments. Aperture-limited arrays reduce the angular-domain diversity available in CSI, thereby weakening LSM’s noise robustness. Therefore, we adopt multi-frequency LSM (MLSM)\footnote{Notably, in practical ISAC systems, the required multi-antenna and multi-frequency observations are inherent to an orthogonal frequency-division multiplexing multiple-input multiple-output (OFDM–MIMO) architecture.}, exploiting frequency diversity to enhance robustness \cite{b15}. Yet, since electromagnetic waves accrue phase differences across frequencies, naive stacking multi-frequency data often leads to destructive interference, thereby producing artifacts in SGR \cite{b16}. To address this, we adopt a wireless-channel perspective on scattered-field propagation. From this perspective, the phase offsets of scattered fields (or CSI) across frequencies stem from propagation through distinct ISAC subchannels. By applying matched filtering to each frequency to enforce inter-frequency phase alignment, destructive interference is turned into constructive interference, improving the robustness of SGR.

\section{System Model Formulation}

\hspace{1em}Consider a two-dimensional scalar inverse scattering problem under transverse-magnetic (TM) polarization. 
Let $D \subseteq \mathbb{R}^2$ denote the domain of interest (DOI), which contains the target scatterer. 
We arrange ${N_\text{Tx}}$ transmitters and ${N_\text{Rx}}$ receivers uniformly along a circumference of radius $R$ in the far-field region $\Gamma$ surrounding $D$.
\begin{figure}[htbp]
    \centering
    \includegraphics[width=0.3\textwidth]{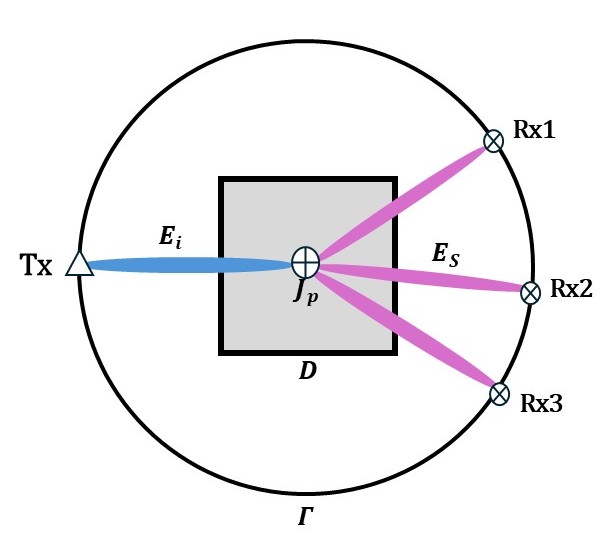} % 调整width控制大小
    \caption{Illustration of LSM.}
    \label{fig:my_label}
\end{figure}

\hspace{1em}Discretizing $D$ into $P$ pixels, at each pixel location $\mathbf{r}_p$, the incident field $\mathbf{E}_i$ induces an equivalent current $\mathbf J_p$. Using superscript $n$ to denote the contribution from the $n$-th transmitter,
the scattered field measured at the receivers $\mathbf{E_{s}^{n}}(\mathbf{R}_{rx})$  can then be expressed using the method of moments (MoM) as
\begin{equation}
\mathbf{E}_{s,k}^{n}(\mathbf{R}_{rx}) = \sum_{p=1}^{P} \mathbf G(\mathbf{R}_{rx}, \mathbf{r}_p) \mathbf J_p^{n},
\end{equation}

where $\mathbf{G}(\mathbf{R}_{rx}, \mathbf{r}_p) = \frac{-j}{4} H_0^{(2)}(k |\mathbf{R}_{rx} - \mathbf{r}_p|)$ $\in \mathbb{C}^{N_{\mathrm{Rx}}\times P}$ is the two-dimensional Green's function from the pixel $\mathbf{r}_p$ to all receivers, 
$H_0^{(2)}$ denotes the zeroth-order Hankel function of the second kind, and $k$ is the wavenumber. 

Arrange all the scattered fields $\mathbf{E}_{s,k}^{n}$ generated by the transmitters 
into a complex matrix $\mathbf{F} \in \mathbb{C}^{N_{\mathrm{Rx}} \times N_{\mathrm{Tx}}}$, 
with each column corresponding to one transmitter:
\begin{equation}
\mathbf{F}_k =
\begin{bmatrix}
\mathbf{E}_{s,k}^{1}(\mathbf{R}_{\mathrm{rx}}) & 
\mathbf{E}_{s,k}^{2}(\mathbf{R}_{\mathrm{rx}}) & 
\cdots & 
\mathbf{E}_{s,k}^{N}(\mathbf{R}_{\mathrm{rx}})
\end{bmatrix},
\end{equation}

the solution of the LSM at $\mathbf{r}_p$ is obtained by solving the following equation at a given frequency $f_k$:
\begin{equation}
\mathbf {F}_k\,\mathbf{g}_k(\mathbf{r}_p) = \mathbf G(\mathbf{R}_{rx}, \mathbf{r}_p).
\end{equation}

\begin{remark}
Equation (3) can be interpreted as an equivalent source problem at pixel $\mathbf{r}_p$ \cite{b17}. If $\mathbf{r}_p$ is located inside the scatterer, an equivalent point current source $\mathbf J_p$ at $\mathbf{r}_p$ will generate the cylindrical scattered wave on the right-hand side of equation (3). Thus, there exists a vector weight $\mathbf{g}_k(\mathbf{r}_p)$ that renormalizes the point current source $\mathbf J_p$ induced by the incident field $ \mathbf E_i$ at this pixel to $\mathbf J_p$. Therefore, equation (3) is equivalent to examining whether an internal point current source $\mathbf J_p$ can exist at pixel $\mathbf{r}_p$, which in turn allows us to determine the positional relationship between $\mathbf{r}_p$ and the scatterer.
\end{remark}

\hspace{1em}Due to the ill-posedness of equation (3), the solution is sensitive to noise and measurement errors. Tikhonov regularization is typically employed:
\begin{equation}
\min_{\mathbf{g}_k} \left\{ \| \mathbf{{F}_k}(\mathbf{R}_{rx}) \,\mathbf{g}_k(\mathbf{r}_p) - \mathbf G(\mathbf{R}_{rx}, \mathbf{r}_p) \|^{2} + \alpha \| \mathbf{g}_k(\mathbf{r}_p) \|^{2} \right\},
\end{equation}
where $\alpha$ is the regularization parameter and $\|\cdot\|$ denotes the Euclidean norm \cite{b18}. By performing singular value decomposition (SVD) on $\mathbf{F}_{k}$, we obtain $\{u_n, \lambda_n, v_n\}$. Then, $\mathbf{g}_k(\mathbf{r}_p)$ can be computed as follows:
\begin{equation}
\mathbf{g}_k(\mathbf{r}_p) = \sum_{n=1}^{\infty} \frac{\lambda_{n}}{\lambda_{n}^{2} + \alpha^{2}} \left\langle \mathbf G(\mathbf{R}_{rx}, \mathbf{r}_p), u_{n} \right\rangle v_{n}.
\end{equation}

\hspace{1em}The single frequency LSM indicator in pixel $\mathbf{r}_p$ is defined as:  $\|\mathbf{g}_k(\mathbf{r}_p)\|^{-2}$, 
and the amplitude contributions of $\mathbf{g}_k$ at different frequencies are summed to obtain the parallel multi-frequency indicator \cite{b15}:
\begin{equation}
\mathbf I_P(\mathbf{r}_p) = \left( \sum_{k=1}^{N_f} \|\mathbf{g}_{k}(\mathbf{r}_p)\|^{-2} \right)^{1/2}.
\end{equation}

%%%%%%%%%%%%%%%%%%%%%%%%%%%%%%%%%%%%%%%%%%%%%%%%%%%%%%%%%

\section{Interpretation of the LSM in Wireless Channels}
\subsection{Viewing LSM System as A Wireless Channel}

\hspace{1em}For notational convenience, in what follows we omit the subscript $k$ and prescript $n$ in this section. By considering only the contribution at a single pixel $\mathbf{r}_p$ in equation (1), and using the field relation $\mathbf{E}_s^{\mathbf{r}_p} = \mathbf{E}_t^{\mathbf{r}_p} - \mathbf{E}_i^{\mathbf{r}_p}$ as well as substituting the equivalent current source expression for $ \mathbf J_p $, we have
\begin{equation}
\mathbf{E}_t^{\mathbf{r}_p}(\mathbf{r'}) - \mathbf{E}_i^{\mathbf{r}_p}(\mathbf{r'}) = \mathbf G(\mathbf{r'}, \mathbf{r}_p) \mathbf \chi(\mathbf{r}_{p}) \mathbf{E}_t^{\mathbf{r}_p }(\mathbf{r}_p),
\end{equation}
where \(\mathbf{E}_t^{\mathbf{r}_p}\) and \(\mathbf{E}_i^{\mathbf{r}_p}\) denote the total and incident fields at \(\mathbf{r}'\) due to the current at \(\mathbf{r}_p\), respectively, with \(\mathbf{r}' \in D\) is an arbitrary point distinct from \(\mathbf{r}_p\).
$\chi(\mathbf{r}_{p})$ is the contrast  at  $\mathbf{r}_p$ defined as:
\begin{equation}
\label{eq:chi}
\chi(\mathbf{r}_p) = \epsilon_r(\mathbf{r}_p) + j\frac{\sigma(\mathbf{r}_p)}{\varepsilon_0 \omega_k } - 1,
\end{equation}
where $\epsilon_r(\mathbf{r}_p)$ is the relative permittivity, $\sigma(\mathbf{r}_p)$ is the conductivity, and $\varepsilon_0$ is the permittivity of free space \cite{b12}. $\omega_k$ is the angular frequency on $k$-th frequency.
Swapping the positions of $\mathbf{E}_i^{\mathbf{r}_p}$ and $\mathbf{E}_t^{\mathbf{r}_p}$ on both sides of the equation and multiplying both sides by $\chi(\mathbf{r}_{p})$, we obtain:
\begin{equation}
(1-\mathbf G(\mathbf{r'}, \mathbf{r}_p)\chi(\mathbf{r}_{p})) \mathbf{J}_{p} =\chi(\mathbf{r}_{p})\mathbf{E}_i^{\mathbf{r}_p}(\mathbf{r_p}).
\end{equation}

\hspace{1em}In the above derivation we assumed $\mathbf{E}_t^{\mathbf{r}_p}$ and $\mathbf{E}_i^{\mathbf{r}_p}$ at $\mathbf{r'}$ and $\mathbf{r_p}$ are approximately equal. This approximation is reasonable under far-field, weak-scattering. Finally we obtained:
\begin{equation}
 \mathbf{J}_{p} =(1-\mathbf G(\mathbf{r'}, \mathbf{r}_p)\chi(\mathbf{r}_{p}))^{-1}\chi(\mathbf{r}_{p})\mathbf{E}_i^{\mathbf{r}_p},
\end{equation}

this is consistent with the conclusion in \cite{b12}. Combining equation (1), (7) and (10), we obtain:

\begin{equation}
\begin{aligned}
\mathbf E_s^{\mathbf{r}_p}(\mathbf{R}_{rx}) &=  \mathbf G(\mathbf{R}_{rx}, \mathbf{r}_p) (1-\mathbf G(\mathbf{r'}, \mathbf{r}_p)\chi(\mathbf{r}_{p}))^{-1}\chi(\mathbf{r}_{p})\mathbf{E}_i^{\mathbf{r}_p}(\mathbf{r_p}) \\
&= \mathbf G(\mathbf{R}_{rx}, \mathbf{r}_p) \mathbf X(\mathbf{r}_p),
\end{aligned}
\end{equation}
$\mathbf X(\mathbf{r}_p)=(1-\mathbf G(\mathbf{r'}, \mathbf{r}_p)\chi(\mathbf{r}_{p}))^{-1}\chi(\mathbf{r}_{p})\mathbf{E}_i^{\mathbf{r}_p}(\mathbf{r_p}) $
represents the ``source signal'' that carries all the undistorted information at $\mathbf{r}_p$.

\hspace{1em}To account for measurement noise, we add complex Gaussian noise to equation (11):
\begin{equation}
\mathbf E_s^{{\mathbf{r}_p}}(\mathbf{R}_{rx}) = \mathbf G(\mathbf{R}_{rx}, \mathbf{r}_p) \mathbf X(\mathbf{r}_p) + \mathbf{n}.
\end{equation}

\hspace{1em}From the above formulation, the scattering process can be interpreted in analogy to a wireless communication channel:
the incident field acts as the transmitted signal, the scatterer generates the source signal $ \mathbf X$, 
which then propagates through the channel characterized by $ \mathbf G$, and additive white Gaussian noise (AWGN) is superimposed at the receiver, resulting in the observed signal $ \mathbf E_s$. 
This perspective enables the application of wireless communication theory and signal processing techniques to the analysis of LSM problems.

\subsection{CSI-Based MLSM with Matched Filtering}

\hspace{1em}In equation (6), the indicator is typically formed by directly summing the amplitudes of the solutions $\mathbf{g}_k$ at different frequencies. 
However, this approach has two main limitations.
\hspace{1em}First, conventional Tikhonov regularization in LSM only constrains the amplitude of the solution and does not exploit the phase information present in the measured data. 
The phase relationships between different frequencies, which can provide valuable information for distinguishing true scatterers from artifacts, are thus ignored. Consequently, 
the indicator function may not fully utilize all available measurement information \cite{b13},
 resulting in suboptimal imaging performance, especially in the presence of noise or closely spaced scatterers.
Second, from the wireless channel perspective in equation (12), 
simply adding the amplitudes of multi-frequency $\mathbf{g}_k$ is equivalent to coherently combining multipath signals that traverse similar propagation paths but have different frequencies. 
Due to the frequency-dependent phase term $\mathbf{\phi = e^{-j 2\pi d f_k / c}}$ (where $\mathbf{d}$ is the distance between $\mathbf{r}_{p}$ and receivers, $\mathbf{f}_{k}$ is the working frequency),
the phases of these signals can vary significantly between frequencies. 
As a result, destructive interference may occur, which can weaken the information about the scatterer carried by the multi-frequency signals. 
This effect can lead to artifacts and reduced image quality.

\hspace{1em}A feasible approach to address the above two challenges is to employ a matched filter:
\begin{equation}
\mathbf S =\mathbf{G}(\mathbf{R}_{rx}, \mathbf{r}_p)^{\mathrm H},
\end{equation}
where the superscript ${\mathrm H}$ denotes the conjugate transpose.

Thus, the received scattered signal can be processed by a matched filter as follows:
\begin{equation}
  \begin{aligned}
\mathbf Y &= \mathbf S \, \mathbf E_{s}^{{\mathbf{r}_p}}(\mathbf{R}_{rx}) 
   = \mathbf{G}(\mathbf{R}_{rx}, \mathbf{r}_p)^{\mathrm H} \, \mathbf E_{s}^{{\mathbf{r}_p}}(\mathbf{R}_{rx})\\
&= \mathbf{G}(\mathbf{R}_{rx}, \mathbf{r}_p)^{\mathrm H}\mathbf{G}(\mathbf{R}_{rx}, \mathbf{r}_p)\, \mathbf X(\mathbf{r}_p)+\mathbf{G}(\mathbf{R}_{rx}, \mathbf{r}_p)^{\mathrm H}\mathbf{n}\\
&=\|\mathbf{G}(\mathbf{R}_{rx}, \mathbf{r}_p)\|^{2}\,\mathbf X(\mathbf{r}_p)+\mathbf{n'}.
\end{aligned}
\end{equation}

\hspace{1em}Since multiplying a matrix by a scalar does not alter its SVD structure, 
the information carried by the matched filter output $\mathbf Y$ is essentially the same as that of the undistorted signal $\mathbf X(\mathbf {r}_p)$, 
except for the presence of AWGN $n'$. By assembling the scattering matrix $\mathbf{F}_k$ using the method introduced in Section~II, 
the LSM can then be performed. 
Employing the matched filter, the potential destructive interference is eliminated. 
Consequently, the scatterer information contained in the multi-frequency signals is fully preserved, thereby enhancing the robustness of the LSM.

\section{Numerical Experiments and Results}
\begin{figure}[htbp]
    \centering
    \includegraphics[width=0.34\textwidth]{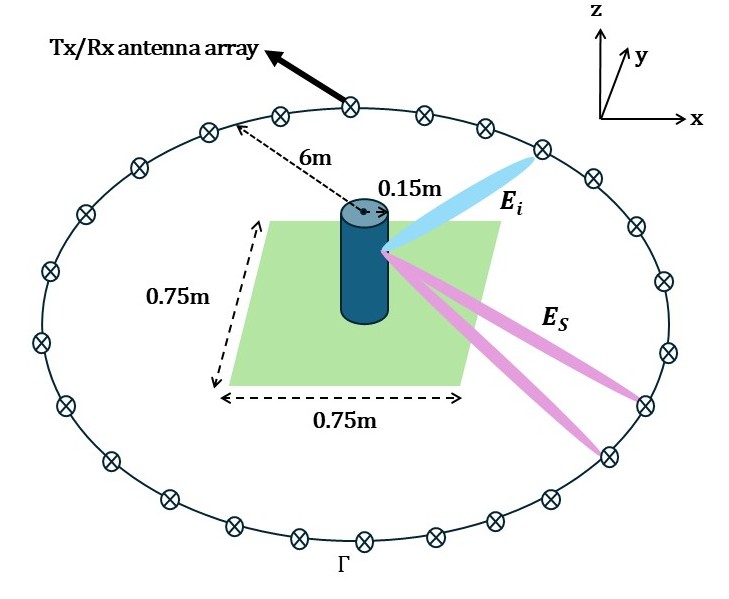} % 调整width控制大小
    \caption{Simplified DOI and the scatterer illustration.}
    \label{fig:my_label}
\end{figure}

\begin{figure}[htbp]
    \centering
    \includegraphics[width=0.4\textwidth]{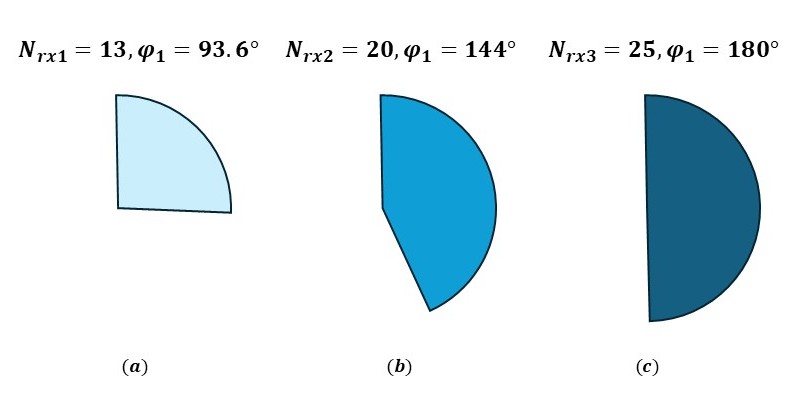} % 调整width控制大小
    \caption{Three limited-aperture receiver array configurations: (a) 13 antennas over $93.6^\circ$, (b) 20 antennas over $144^\circ$, and (c) 25 antennas over $180^\circ$.}
    \label{fig:my_label}
\end{figure}

\hspace{1em}In the rest of this letter, we refer to matched-filtering–enhanced MLSM as MF-MLSM. In the following experiments, a $0.75\mathrm{m}\times0.75\mathrm{m}$ square centered at the origin is defined as the DOI (Fig.~2). The DOI size is chosen to fully enclose the scatterer with a safety margin while keeping the inversion grid and runtime manageable. We model the scatterer as a $z$-invariant circular cylinder of radius $0.15\mathrm{m}$. While practical environments often contain multiple or extended scatterers, we adopt a single target to validate the method and avoid multi-object interactions. Under TM$_z$ polarization, the electric field is $z$-independent, yielding a 2D model. SGR is equivalently to reconstruct the boundary of a circular scatterer of radius $0.15\mathrm{m}$ within the square DOI. The DOI contains 2{,}116 pixels, 266 in the circular region. On a circle of radius $6\mathrm{m}$ (far field), $50$ antennas are distributed. Each antenna acts as transmitter and receiver.

\hspace{1em}Eight frequencies (from $1$ to $8\,~\mathrm{GHz}$ in $1\,~\mathrm{GHz}$ steps) are used to generate scattered field data. 
To assess noise robustness, AWGN is injected at 27~dB and 12~dB. 
We then compare MLSM and MF\mbox{-}MLSM across all eight frequencies under both noisy and limited\mbox{-}aperture settings. 
Receiver sparsity is evaluated with three limited\mbox{-}angle arrays as shown in Fig.~3: 13 antennas over $93.6^\circ$, 20 antennas over $144^\circ$, and 25 antennas over $180^\circ$.

 \hspace{1em}Figs.~4 and 5 present the reconstructions of MLSM and MF-MLSM under $180^\circ$ and $144^\circ$ limited-aperture scenarios at two SNR levels. For each configuration, the Tikhonov regularization parameter \(\alpha\) is selected via the L-Curve method \cite{b19}. With 27~dB SNR, MF-MLSM reconstructs the boundary more accurately than MLSM on both $180^\circ$ and $144^\circ$ limited, as evidenced by comparing Fig.~4(a) with Fig.~4(c) and Fig.~5(a) with Fig.~5(c). While MLSM recovers a roughly circular region, noticeable boundary underfill remains, consistent with destructive multi-frequency interference caused by phase jumps. In contrast, MF-MLSM eliminates inter-frequency phase discontinuities through matched filtering, enabling constructive combination of scattering information and yielding a more complete boundary. Quantitatively, Table~\ref{tab:recon_pixels} shows the reconstructed area coverage increases from 70.68\% to 98.12\% for the 180$^\circ$ aperture and from 73.78\% to 97.74\% for the 144$^\circ$ aperture. All percentages are computed as the ratio of reconstructed pixels to the 266-pixel full-circle ground-truth region. At 12~dB SNR, both methods exhibit varying degrees of distortion (Figs.~4(b), ~4(d) and ~5(b), ~5(d)). However, according to Table~\ref{tab:recon_pixels}, MF-MLSM attains a higher reconstructed area, remaining above 94\% under both angular aperture limits. MF-MLSM also exhibits lower distortion than MLSM under these conditions. Overall, MF-MLSM consistently achieves higher area coverage and cleaner, less-distorted boundaries than MLSM across apertures and SNRs, demonstrating superior robustness under noisy, limited-aperture conditions.

\begin{table}[htbp]
  \centering
  \caption{Reconstructed area as a percentage of the ground-truth scatterer (266 pixels).}
  \label{tab:recon_pixels}
  \setlength{\tabcolsep}{6pt}
  \renewcommand{\arraystretch}{1.1}
  \begin{tabular}{@{}lcc@{}}
    \toprule
    & \multicolumn{2}{c}{Reconstructed area relative to ground truth } \\
    \cmidrule(lr){2-3}
         Setting  & MLSM(\%) & MF-MLSM(\%) \\
    \midrule
    Aperture $180^\circ$, SNR = 27~dB & 70.68 & 98.12 \\
    Aperture $180^\circ$, SNR = 12~dB & 67.30 & 94.00 \\
    Aperture $144^\circ$, SNR = 27~dB & 73.78 & 97.74 \\
    Aperture $144^\circ$, SNR = 12~dB & 71.92 & 96.62 \\
    \bottomrule
  \end{tabular}
\end{table}

\FloatBarrier

\hspace{1em}Fig.~6 shows the imaging results for MLSM and MF-MLSM at $93.6^\circ$ aperture and 27~dB SNR. Under this strong aperture limitation, both algorithms suffer from significant and similar boundary distortions. This indicates that the coherent gain provided by matched filtering is insufficient to compensate for the increased ill-posedness of equation (3) caused by the lack of angular information. Under such conditions, MF-MLSM does not offer a clear advantage over MLSM.

%%%%%%%%%%%%%%
\begin{figure}[htbp]
    \centering
    \includegraphics[width=0.5\textwidth]{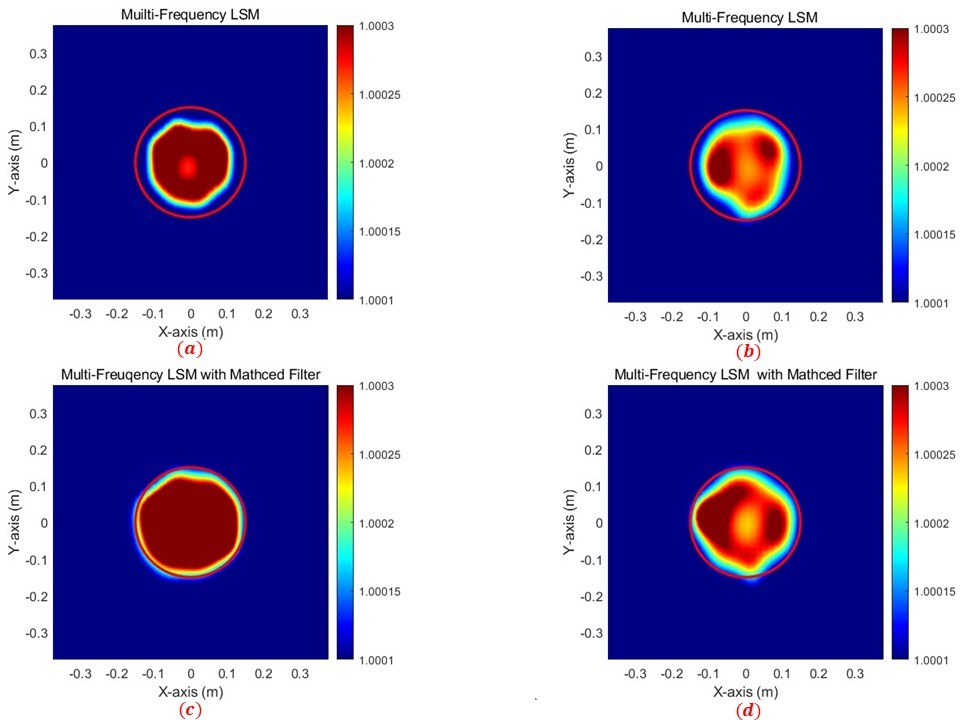} 
    \caption{Aperture over $180^\circ$. (a) MLSM with SNR=27~dB, $\mathbf{\alpha=10^{-2}}$; (b) MLSM with SNR=12~dB, $\mathbf{\alpha=10^{-6}}$; (c) MF-MLSM with SNR=27~dB, $\mathbf{\alpha=10^{-6}}$; (d) MF-MLSM with SNR=12~dB, $\mathbf{\alpha=10^{-6}}$.}
    \label{fig:my_label}
\end{figure}

%%%%%%%%%%%%%
\begin{figure}[htbp]
    \centering
    \includegraphics[width=0.5\textwidth]{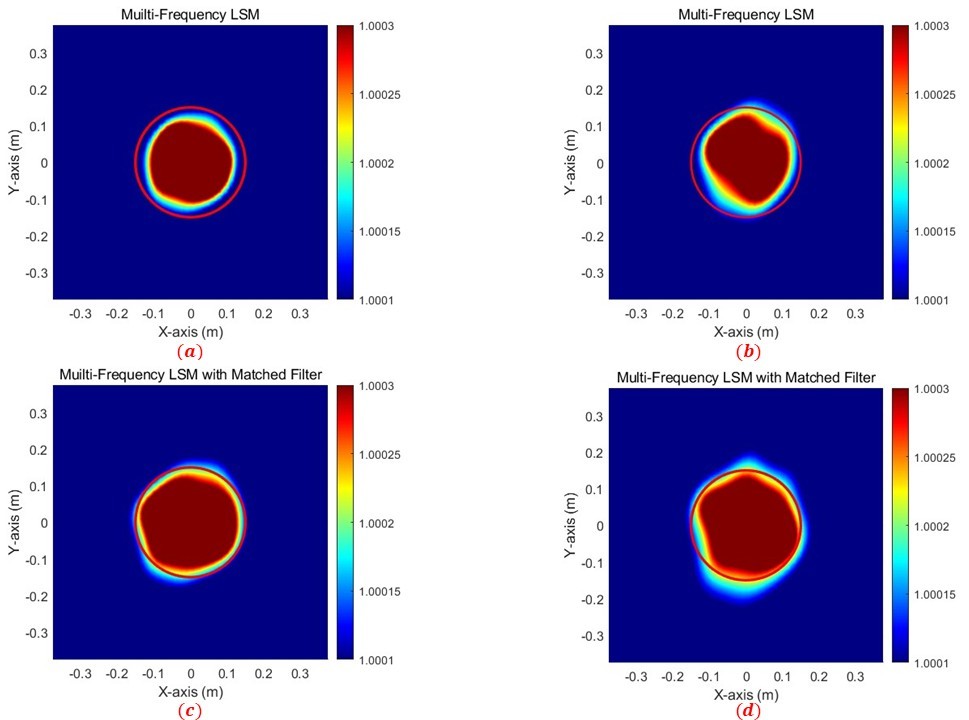} % 调整width控制大小
    \caption{Aperture over $144^\circ$. (a) MLSM with SNR=27~dB, $\mathbf{\alpha = 1 \times 10^{-2}}$; (b) MLSM with SNR=12~dB, $\mathbf{\alpha = 1 \times 10^{-6}}$;
    (c) MF-MLSM with SNR=27~dB, $\mathbf{\alpha = 1 \times 10^{-6}}$; (d) MF-MLSM with SNR=12~dB, $\mathbf{\alpha = 1 \times 10^{-6}}$.}
    \label{fig:my_label}
\end{figure}
\begin{figure}[htbp]
    \centering
    \includegraphics[width=0.5\textwidth]{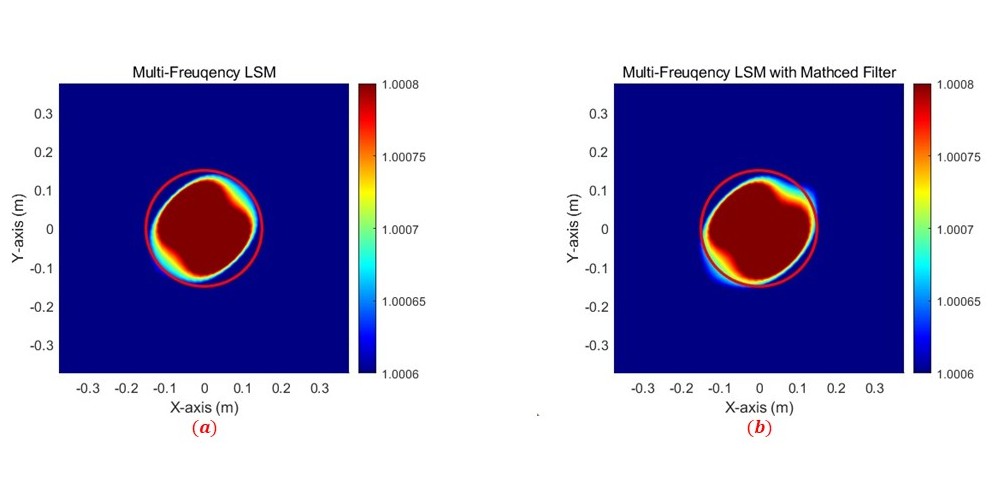} % 调整width控制大小
    \caption{Aperture over $93.6^\circ$. (a) MLSM with SNR=27~dB, $\mathbf{\alpha = 1 \times 10^{-2}}$; (b) MF-MLSM with SNR=27~dB, $\mathbf{\alpha = 1 \times 10^{-6}}$.}
    \label{fig:my_label}
\end{figure}

\section{Conclusion}

\hspace{1em}In this letter we integrate scatter geometric reconstruction in ISAC system with wireless channel theory and reinterpret LSM from a channel perspective. To counter phase jumps across frequencies, we adopt matched filtering. Simulations show matched-filtered MLSM provides superior reconstruction under limited apertures (93.6$^\circ$, 144$^\circ$, 180$^\circ$) and noisy conditions (27 and 12~dB SNR). 

\newpage

% Can use something like this to put references on a page
% by themselves when using endfloat and the captionsoff option.

\end{document}